Electronic structure-property relationship in an Al0.5TiZrPdCuNi high-entropy alloy

Emil Babić[a]*, Ignacio A. Figueroa[b], Vesna Mikšić Trontl[c], Petar Pervan[c], Ivo Pletikosić[d,e], Ramir Ristić[f], Amra Salčinović Fetić[g]**, Željko Skoko[a], Damir Starešinić, Tonica Valla[d,h], Krešo Zadro[a]

[a] Department of Physics, Faculty of Science, Bijenička Cesta 32, HR-10002, Zagreb, Croatia

[b] Institute for Materials Research-UNAM, Ciudad Universitaria Coyoacan, C.P. 04510, Mexico D.F., Mexico

[c] Institute of Physics, Bijenička Cesta 46, HR-10000 Zagreb, Croatia

[d] Condensed Matter Physics and Materials Science Department, Brookhaven National Lab, Upton, NY 11973, USA;

[e] Department of Physics, Princeton University, Princeton, NJ 08544, USA

[f] Department of Physics, University of Osijek, Trg Ljudevita Gaja 6, HR-31000 Osijek, Croatia

[g] University of Sarajevo, Faculty of Science, Zmaja od Bosne 35, 71000 Sarajevo, Bosnia and Herzegovina

[h] Donostia International Physics Center, 20018 Donostia-San Sebastian, Spain

Corresponding authors:

*E-mail address: ebabic@phy.hr (E. Babić), **E-mail address: amra.s@pmf.unsa.ba

Abstract

The valence band (VB) structure of an Al0.5TiZrPdCuNi high-entropy alloy (HEA) obtained from X-ray photoelectron spectroscopy has been compared to that recently calculated by Odbadrakh et al, 2019. Both experimental and theoretical VBs show split-band structures typical of alloys composed from the early (TE) and late (TL) transition metals. Accordingly, several properties of this alloy (both in the glassy and crystalline state) associated with the electronic structure (ES), are compared with those of similar TE-TL alloys. The comparison shows in addition to the usual dependence on the total TL content strong effect of alloying with Al on the density of states at the Fermi level, $N(E_F)$ and on the magnetic susceptibility of Al0.5TiZrPdCuNi HEA, which is like that of conventional glassy alloys, such as Zr-Cu-Al ones. Despite some similarity between the shapes of theoretical and corresponding experimental VBs there are significant quantitative differences between them which should be taken into account in any future studies of ES in HEAs and other compositionally complex alloys (CCA).

The multi-principal solid solutions such as nearly equimolar HEAs[1-6] and other CCAs[7,8] explore the middle section of the multicomponent phase diagrams which makes an enormous number of new alloys and compounds available for research and possible applications.[5-7,9-11] As a result of the huge research efforts, large progress has been made in the knowledge and sometimes understanding of HEAs and CCAs. Moreover, some technologically relevant alloys, such as those with excellent mechanical properties as well as good irradiation, corrosion, and oxidation resistance, have been discovered.[5,10,11]

The largest problem in the discovery of CCA systems possessing desirable properties is still limited understanding of these complex systems, which is mostly due to the lack of detailed insight into their ESs. Indeed, ES determines all intrinsic properties of metallic systems including the formation of alloys, their stability, and all physical properties.[5,12-24] Because of this, the number of theoretical studies of ESs and selected properties (mostly mechanical ones) of HEAs and CCAs increased dramatically over the last years and at present, largely exceeds the number of experimental studies of their ES.

However, as noted in our previous reports,[8,25-28] a combination of experimental results from the photoemission spectroscopy (PES) and low temperature specific heat (LTSH) with theoretical calculations is required to obtain reliable and quantitative insight into the ES of the studied alloy. Indeed, a detailed insight into the ES and superconducting and other properties of binary TE-TL glassy alloys, such as Zr-TL alloys (TL=Fe, Co, Ni, Cu, Rh and Pd) has been obtained through comparison of the calculated VB structure with that obtained from PES.[8,29] Unfortunately, there is a small number of PES studies performed on HEAs and CCAs[8,25-28,30,31] and there is no comparison between the experimental and theoretical electronic density of states (DOS) within VB for these alloy systems.

The recent X-ray photoemission spectra (XPS) for Al0.5TiZrPdCuNi glassy alloy[8] enabled the first comparison between the theoretical[32] and experimental VB structure of an HEA. This alloy is important because it can be prepared in both amorphous and crystalline state (with a dominant body centred cubic (bcc) structure).[33-35] Thus, the study of this alloy and other CCAs[28,36-38] and conventional alloys[19,20,39,40] sharing this property could help to disentangle the effect of topological and chemical disorder on their properties.

Accordingly, we prepared[8] both amorphous and crystalline samples of Al0.5TiZrPdCuNi alloy and in addition to XPS performed the magnetic susceptibility and microhardness measurements on two types of samples. Both, experimental (XPS) and theoretical DOS revealed a split-band shape of VB with the d-states of TEs (Ti and Zr) dominant in DOS close to the Fermi level, $E_F$ which is usual for TE-TL alloys.[8,27,29,41,42] Thus, $N(E_F)$[35] and all associated properties depend for a given TEs on a total TL content only[43] (here, Pd+Cu+Ni), while alloying with Al causes an additional decrease of these properties. The first direct comparison between the experimental and theoretical VB structure despite the qualitative similarity, shows considerable quantitative differences between them. This emphasizes the importance of the combined experimental and theoretical studies of the ES of HEAs and CCAs.

The method of preparation of Al0.5TiZrPdCuNi samples has been reported in Ref. 8. Both the melt-spun ribbons and a cylinder with a diameter of 1.5 mm obtained by the mold-casting technique were prepared to verify a claim[33] that this alloy can be prepared in either an amorphous or single-phase bcc crystalline structure, depending on the cooling rate from the melt. However, as seen from Fig. S1 in the supplementary material as-cast ribbons were amorphous, but the cylinder had three phase structure, like these obtained in other attempts[34,35] to obtain a single-phase bcc structure. The lattice parameter of a dominant (about 70%) bcc phase, a=3.10 A (see supplementary material for full discussion of X-ray diffraction, XRD results) was the same as that obtained in[35] and was also close to these reported in Ref.33 and 34, a=3.125 A.[32] For a bcc lattice this lattice parameter implies the average atomic radius of $r_{ave}$ =1.35 A which is well below of that obtained from the Vegard law, $r_V$=1.40 A.[33] Detailed characterization of these samples (see supplementary material) included the thermal analysis, DSC/DTA (Fig.S2), which confirmed the amorphous state of the as-cast ribbons and provided the values of thermal parameters and the SEM-EDS analysis (Fig S3) of their actual composition and the chemical homogeneity. The chemical composition results agreed well with nominal composition, except for some excess of Cu and deficiency in Ni content and the distribution of elements appeared random down to a micrometre scale (Fig. S3). The methods of the measurements performed on these samples: XPS, magnetic susceptibility and microhardness, were described in some detail in our previous reports.[8,28,31]

The comparison of the theoretical VB structure (DOS, Ref. 32) with that from the XPS spectrum for Al0.5TiZrPdCuNi alloy in Fig. 1 is important result of this research. The theoretical ES of this alloy in the bcc lattice was calculated by using Korringa-Kohn-Rostoker coherent-potential-approximation (KKR-CPA) within the generalized gradient approximation for exchange-correlation potential.[32] The supercell consisting of 256 atoms was created using the KKR-CPA optimized lattice parameter, a=3.115 A, where the sites of a perfect bcc lattice are populated randomly with constituent atoms without any chemical short-range order (SRO). The XPS spectrum, associated with the variation of DOS within the VB was measured on amorphous ribbon,[8] but these spectra are practically the same for amorphous and corresponding crystallized samples, both in HEAs[28] and conventional alloys.[29] However, the XPS spectrum in addition to DOS reflects the variation of the photoionization cross section, σ which depends on the type of atom and the electron orbital (s, p, d, f) as well as the energy/frequency of employed photons. (In transition metal alloys d-electrons have the largest σ.[8]) Thus, the XPS spectrum must be corrected to obtain the exact DOS. This correction requires a detailed insight into the ES of studied alloy which has not been made available to us by the authors of Ref.32. Accordingly, we will only provide a qualitative explanation of the difference between the two representations of DOS in Fig. 1, which is based on the values of σ of constituents, and previous XPS result for similar TiZrNbCuNi alloy (Fig.2 in Ref.8) which is almost the same as that shown in Fig. 1.

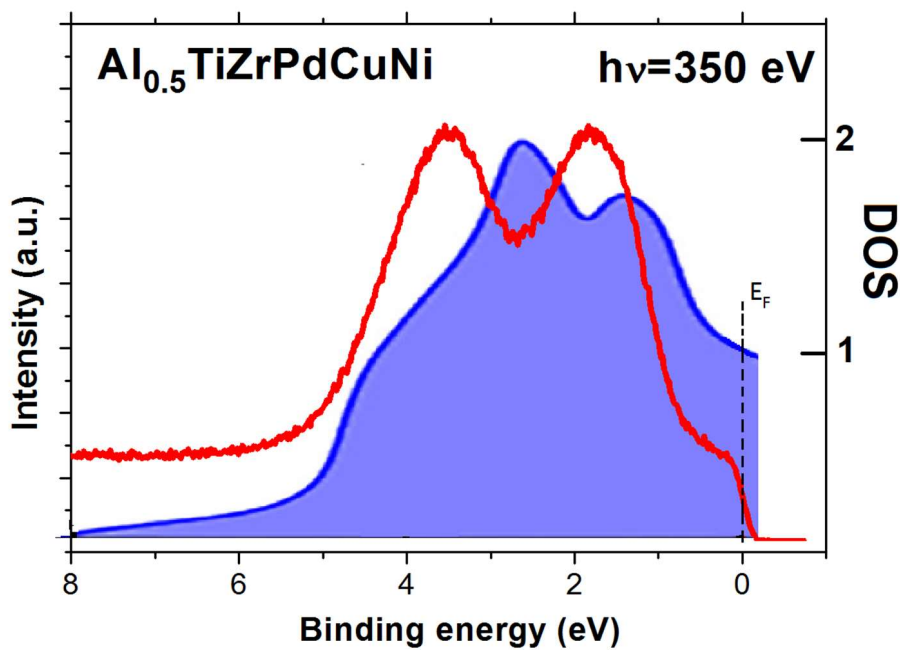

Fig. 1: XPS spectrum (red curve, Ref. 8) and calculated DOS (blue, Ref. 32) for Al0.5TiZrPdCuNi high entropy alloy.

Both, calculated DOS and XPS spectrum show a typical split-band structure of VB with a knee close to $E_F$ followed with two pronounced maxima at higher binding energies, $E_B$. However, the size of the knee and the positions and shapes of maxima at higher $E_B$ are very different in two representations of DOS in Fig.1. A low knee in the XPS spectrum compared to that in DOS is possibly due to a combination of rather small σ of d-electrons of TE= Ti and Zr (relative to these of Ni and Cu) and a rather small content of TEs in Al0.5TiZrPdCuNi, but the alloying with Al may also decrease N($E_F$) as shown in Figs. 2 and 3. Indeed, a knee in XPS of TiZrNbCuNi alloy having higher TE content was about 50% higher than that in Fig.1. Similarly, a deeper minimum between two maxima in the XPS spectrum than that in theoretical DOS, may arise from a combination of a very broad 4d-band of Pd[44,45] and lower σ of 4d states of Pd compared to those of 3d-states of Cu and Ni. This minimum was even deeper in the XPS of TiZrNbCuNi alloy[8] which does not contain Pd but has very similar Cu and Ni contents to these in Al0.5TiZrPdCuNi alloy. Here the cross section for 4d-states of Nb is about 3 times smaller than that for Pd. However, the positions of two maxima in the XPS spectra of both TiZrNbCuNi and Al0.5TiZrPdCuNi alloy are at practically the same binding energies of about 1.85 and 3.55 eV respectively, whereas these in theoretical DOS are centred closer to $E_F$ around $E_B$=1.35 and 2.6 eV respectively. The maxima in the XPS spectrum of TiZrNbCuNi alloy were unambiguously assigned (by varying the Ni[26,30] or Cu[27] content) to the dominance of 3d-states of Ni (lower $E_B$) and Cu (higher $E_B$). Accordingly, two maxima of spectral intensity in Al0.5TiZrPdCuNi probably reflect the centres of the same 3d-states (superimposed on a broad 4d band of Pd) and thus their strong shift towards $E_F$, predicted by the theoretical DOS[32] does not seem likely. Alloying of transition metals and alloys with Al has strong effect on their ES (Fig. 2). The XPS study of PdAl alloy[45] has shown that the alloying of Pd with Al shifts the spectral maximum associated to 4d states of Pd to higher $E_B$ relative to that of a pure Pd and thus decreases N($E_F$) of the alloy. Taking all these findings into account, a low binding energy maximum of spectral intensity in Al0.5TiZrPdCuNi reflects a non trivial interplay of contributions from Ni-3d and Pd-4d bands influenced by alloying with Al and other constituents.

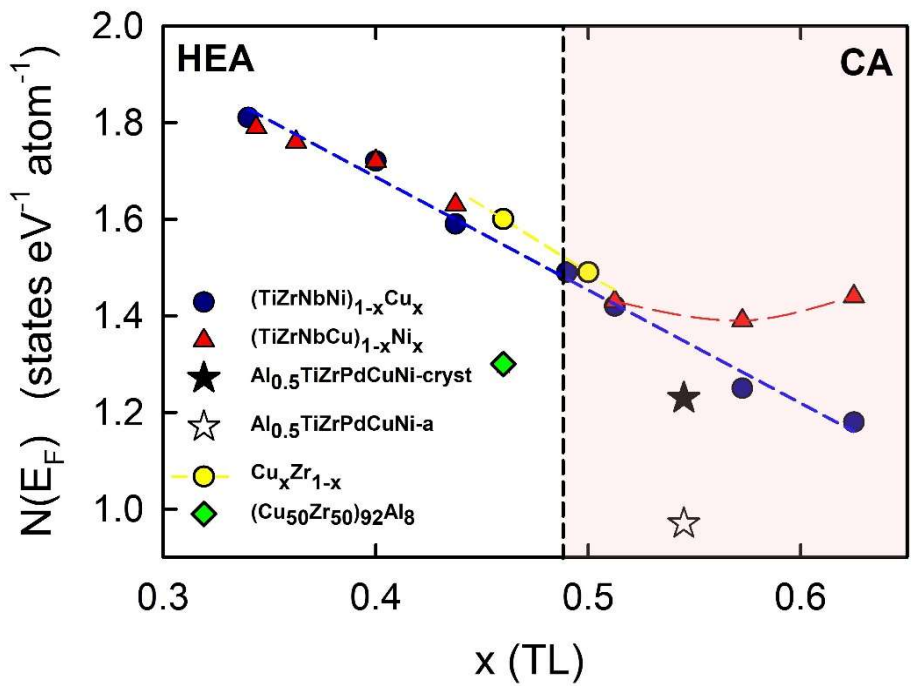

Fig. 2: Density of states at the Fermi level of glassy and crystalline Al0.5TiZrPdCuNi alloy,[35] Ti-Zr-Nb-Cu-Ni alloys[26,28] and Al-doped[47] and undoped[42] Cu-Zr alloys vs. total content of late transition metals, x(TL). Dashed lines are guide to the eyes.

At present we have no proper explanation for this discrepancy, except for the assumption that approximations used in the calculations place the d-bands positions closer to Fermi level. In addition, the assumption of no chemical SRO used in the calculations[32] seems to us questionable considering very strong interactions between the TE and TL atoms[23,26,28,31] and even stronger interactions between these atoms and Al ones. Indeed, recent calculations of DOS for the Cantor alloy have shown very strong influence of chemical SRO on the DOS.[46] This occurs because the dominant structure in DOS obtained from the energy-band calculations is due primarily to the nearest-neighbour chemical environment of each atom.[29]

The knowledge of $N(E_F)$ is necessary to verify whether the theory properly describes ES of a studied system or not.[29] Since there is no quantitative description of the theoretical DOS of Al0.5TiZrPdCuNi alloy in Ref. 32 we will compare recent experimental results for $N(E_F)$ for this alloy (both in a glassy and crystalline state[35]) with these for similar glassy HEA[26,27,30] and conventional TE-TL glassy alloys.[42-47] As already noted in the nonmagnetic TE-TL alloys and in the absence of the band-crossing (the transition from TE dominated to TL dominated $N(E_F)$ on increasing TL content [8,48]) $N(E_F)$ is a nearly universal function of a total TL content, $x(TL)$.[26,28,43] Thus, in Fig. 2 we compare the $N(E_F)$s for amorphous and crystalline Al0.5TiZrPdCuNi alloys[35] with these for $(TiZrNbCu)_{1-x}Ni_x$ and $(TiZrNbNi)_{1-x}Cu_x$ glassy alloys.[26,27,28,30,41] The $N(E_F)$ of both crystalline and amorphous sample is lower than these of Ti-Zr-Nb-Cu-Ni alloys with the same $x(TL)=0.545$. Since the crystalline sample is composed from three phases (Fig. S1 and corresponding text) with unknown individual contributions to LTSH we will not discuss his $N(E_F)$ any further. (We note however that fairly large $N(E_F)$ of crystalline sample seems to support theoretical prediction of very high $N(E_F)$ of Ti and Zr in a bcc phase.[49]) The $N(E_F)$ of glassy sample is about 28% lower than that of $(TiZrNbNi)_{1-x}Cu_x$ glassy alloy with equivalent $x(TL)$ and over 30% lower than that of the corresponding $(TiZrNbCu)_{1-x}Ni_x$ alloy. The later alloys show the band-crossing for $x(TL) \geq 0.57$ (Fig. 2 and Ref. 30) when the centroid of 3d states of Ni reaches $E_B \leq 1.79$ eV.[8] Accordingly, the position of the first maximum in the calculated DOS at $E_B=1.35$ eV (Fig. 1) would probably imply the band-crossing, thus enhanced $N(E_F)$ which is not observed (Fig. 2). Very low $N(E_F)$ of glassy alloy is probably the result of alloying with Al which is known to reduce the $N(E_F)$ in both glassy[47] and crystalline[45] transition metal alloys. Indeed, as seen in Fig. 2 $N(E_F)$ of glassy $Zr_{0.46}Cu_{0.46}Al_{0.08}$ alloy[47] is about 20% lower than that of the corresponding Al-free glassy alloy. Rather low $N(E_F)$ probably explains the observed absence of superconductivity[30] in the studied alloy.[35] We note that lower $N(E_F)$ is usually associated with a greater stability of an alloy. Although an interpretation of rather low $N(E_F)$ of glassy Al0.5TiZrPdCuNi alloy in terms of the effect of alloying with Al on its ES seems plausible, the measurement of LTSH (and possibly of XPS spectrum) of the quinary TiZrPdCuNi glassy alloy[33] is required to verify this claim.

The magnetic susceptibility, $\chi$ of the normal metals and their alloys is proportional to $N(E_F)$, but in the transition metal ones the description of $\chi$ requires three major terms and of these terms only the term describing the paramagnetic susceptibility of d-electrons enhanced by the Stoner factor is related to $N(E_F)$.[26,27,42,50] Still, the variation of $\chi$ with composition in all glassy TE-TL alloys is qualitatively the same as that of $N(E_F)$, as illustrated in Fig.3 for the alloy systems shown in Fig. 2. Indeed, Fig. 3 is qualitatively the same as Fig. 2 and the main difference is that the suppression of $\chi$ on alloying with Al is in both our Zr-Cu-Al glassy alloys and in Al0.5TiZrPdCuNi ones larger than that of the corresponding $N(E_F)$s, about 40%. This may imply that the partially covalent bonding of Al atoms with transition metal ones strongly reduces the overall paramagnetic behaviour of such alloys (possibly enhancing the diamagnetic contributions to total $\chi$). The other difference between the behaviours of $\chi$ and $N(E_F)$ shown in Fig.3 and Fig. 2 respectively, is a small difference between the values of $\chi$ of crystalline and glassy Al0.5TiZrPdCuNi alloy. This may imply that the effects of alloying with Al are similar in a glassy and the corresponding crystalline alloy,[45] but may also be partially due to a large orbital paramagnetic contribution[26,27,42,50] to $\chi$ of transition metal alloys which is rather insensitive of the actual atomic arrangements (glassy or crystalline) in a given alloy.[19,20] However, the three-phase structure of the crystalline sample makes the discussion of the magnitude of its χ unreliable.

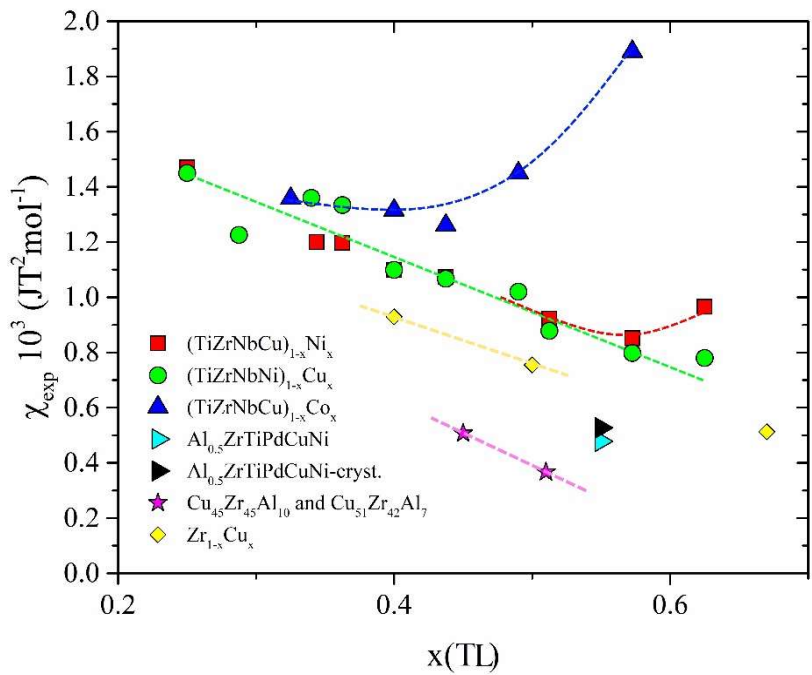

Fig. 3: Room temperature magnetic susceptibility of alloys shown in Fig. 2 and Ti-Zr-Nb-Cu-Co glassy alloys[31] vs. x(TL). Dashed lines are guide to the eyes.

Due to rather simple split-band structure of VB of TE-TL glassy alloys there is a simple correlation between several of their properties, such as the thermal, magnetic, mechanical (including hardness), and electronic (including superconducting) properties and ES. This correlation first observed in the binary TE-TL glassy alloys[51] also applies to HEA type of TE-TL glassy alloys,[26,28] despite of their chemical complexity. Initially we believed that a simple correlation between Hv and the total TL content X(TL) (and thus also with $N(E_F)$, Fig.2) is specific to glassy TE-TL alloys and uncommon in crystalline alloys.[51] However, a recent study of the crystalline structures and the yield strength, $\sigma_y$ of $(TiZrHf)_x(CuNi)_{1-x}$ crystalline alloys (thereafter Ti-Zr-Hf-Cu-Ni) showed that the $\sigma_y$ decreases with increasing x (and thus the TE content) although the crystalline phases also change with x.[52] This implies that the effect of ES on the strength of these quinary TE-TL crystalline alloys is stronger than that of the corresponding crystalline phase(s). In Fig. 4 we compare the microhardness, Hv of Al0.5TiZrPdCuNi samples with these in glassy $(TiZrNbCu)_{1-x}Ni_x$,[26] $(TiZrNbNi)_{1-x}Cu_x$,[28] $(TiZrNbCu)_{1-x}Co_x$,[31] $Cu_{0.45}Zr_{0.55}$ and $Cu_{0.45}Zr_{0.45}Al_{0.1}$ alloys. The variation of $\sigma_y$ of crystalline Ti-Zr-Hf-Cu-Ni alloys[52] with the total TL content, x(TL) is also shown in Fig.4.

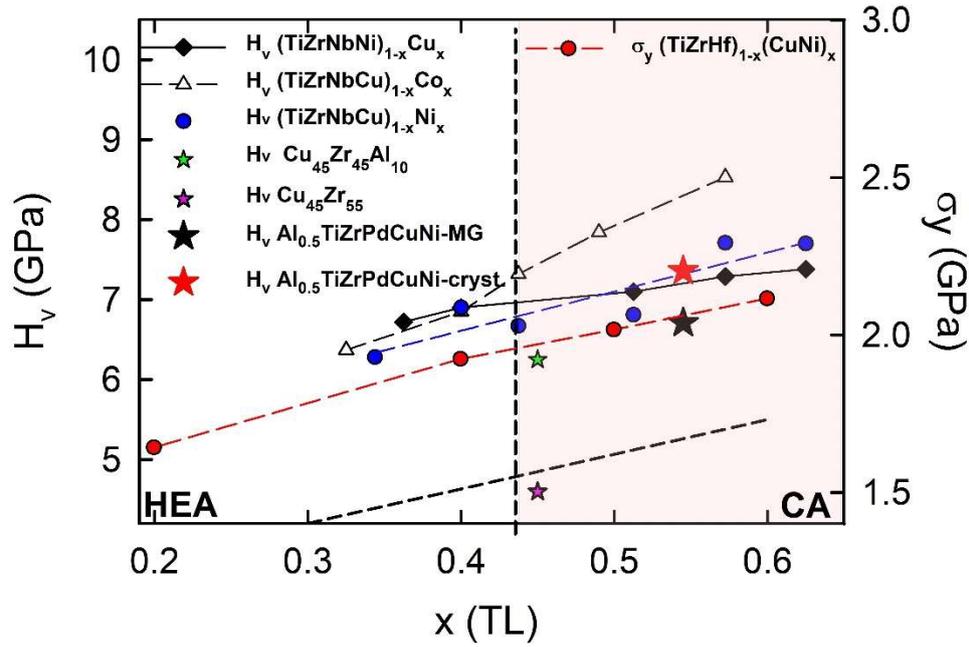

Fig. 4: Microhardness of the alloys shown in Fig. 3 and yield strength of crystalline $(TiZrHf)_{1-x}(CuNi)_x$ alloys[52] vs. x(TL). Dashed lines are guide to the eyes.

The microhardness of a glassy Al0.5TiZrPdCuNi alloy, Hv=6.65 GPa is some 9% lower than that of glassy Ti-Zr-Nb-Ni-Cu alloys with the same x(TL), hence it does not seem to show strong enhancement of Hv caused by alloying with Al observed in glassy $Zr_{0.45}Cu_{0.45}Al_{0.10}$ alloy (note that the strong enhancement of Hv of Al-doped Zr-Cu alloy in Fig. 4 is accompanied with a similarly large decrease of $N(E_F)$ shown in Fig. 2). We note however that Hv of glassy TE-TL is very sensitive to a type of TE(s): the alloys with TE=Ti and Nb have higher Hv that these with TE=Zr with the same TL content.[28] Thus, the enhancement of Hv due to rather large Nb content in the Ti-Zr Nb-Ni-Cu alloys could be larger than that caused by 8% Al in Al0.5TiZrPdCuNi alloys. Somewhat higher Hv of the crystalline Al0.5TiZrPdCuNI sample compared to that of a glassy one is probably due to nanoprecipitates of the phase of type of $Ni_{10}Zr_7$.[35] Since there is the proportionality between Hv, $\sigma_y$ and the Youngs modulus, E in glassy alloys[26,28,31,42,51] the knowledge of one of these parameters is sufficient to determine the values of the other two. From Hv=6.65 GPa for glassy Al0.5TiZrPdCuNi alloy we obtain its $\sigma_y$=Hv/3=2.23 GPa and E=15Hv=100 GPa. All these parameters are about 10% lower than these of Ti-Zr-Nb-Cu-Ni alloys with the same x(TL). However, the estimated $\sigma_y$ of glassy Al0.5TiZrPdCuNi alloy is some 10% larger than that of crystalline Ti-Zr-Hf-Cu-Ni alloy with a similar x(TL).

In summary we note that the comparison of calculated DOS for Al0.5TiZrPdCuNi with that inferred from photoemission spectroscopy revealed considerable difference between them. This difference is unlikely to result from the differences in photoionization cross section of the constituent elements only. This emphasizes the importance of the experimental verification of the theoretical calculations of the electronic structure of compositionally complex alloys. The photoemission spectroscopy results combined with these from the low temperature specific heat[35] provided good insight into the trend of selected physical properties of Al0.5TiZrPdCuNi alloy as compared to those of similar disordered compositionally complex alloys composed from the early and late transition metals. A rather small content of aluminium in this alloy strongly affects the properties which are associated with the electronic density of states at the Fermi level, such as the magnetic susceptibility and superconductivity.

Supplementary material section

X-ray diffraction

Powder X-ray Diffraction (PXRD) data were collected using Bruker Advance D8 diffractometer equipped with a LYNXEYE XE-T detector (Karlsruhe, Germany). Measurements were taken in Bragg-Brentano geometry (1D) with Cu$K\alpha$ radiation (1.54 Å) in the angular range $2\Theta$ from 20-94° with a step size of 0.015° and measuring time of 1s/step.

The XRD pattern of as-cast ribbon is typical for amorphous structure exhibiting broad halo centred at $2\Theta \approx 40°$ (Fig. S1a). On the other hand, the XRD pattern of the cylinder sample (Fig.S1b) shows three crystalline phases - two major phases and a one minor. The two major phases consist of (1) a body-centred cubic (bcc) phase characterized by a unit cell parameter of $a$ = 3.095 Å, and (2) a face-centred cubic (fcc) phase, space group $Fm\overline{3}m$, $Pd_2TiAl$-type (Heusler alloy). Unit cell parameter of the fcc phase is $a$ = 6.20 Å (notably, this value is precisely double that of the bcc phase, as discussed in Ref. 35. Meanwhile, the minor phase is identified as $Ni_{10}Zr_7$-type with an orthorhombic structure, belonging to space group $Aea2$ and lattice parameters $a$ = 9.230 Å, $b$ = 9.178 Å, and $c$ = 12.340 Å. The estimated amount of minor $Ni_{10}Zr_7$-type phase is less than 10 wt%, whereas the two major phases, bcc phase an Pd2TiAl-type phase account for roughly 65 wt% and 25 wt%, respectively, of the sample.

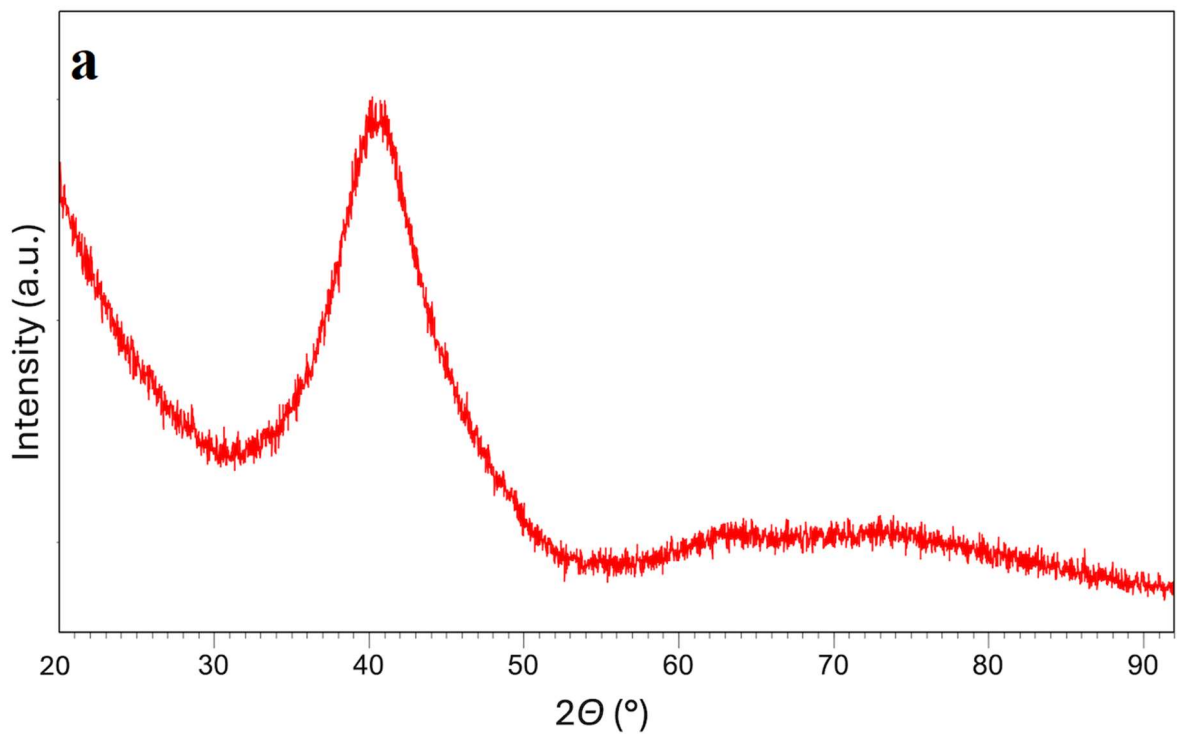

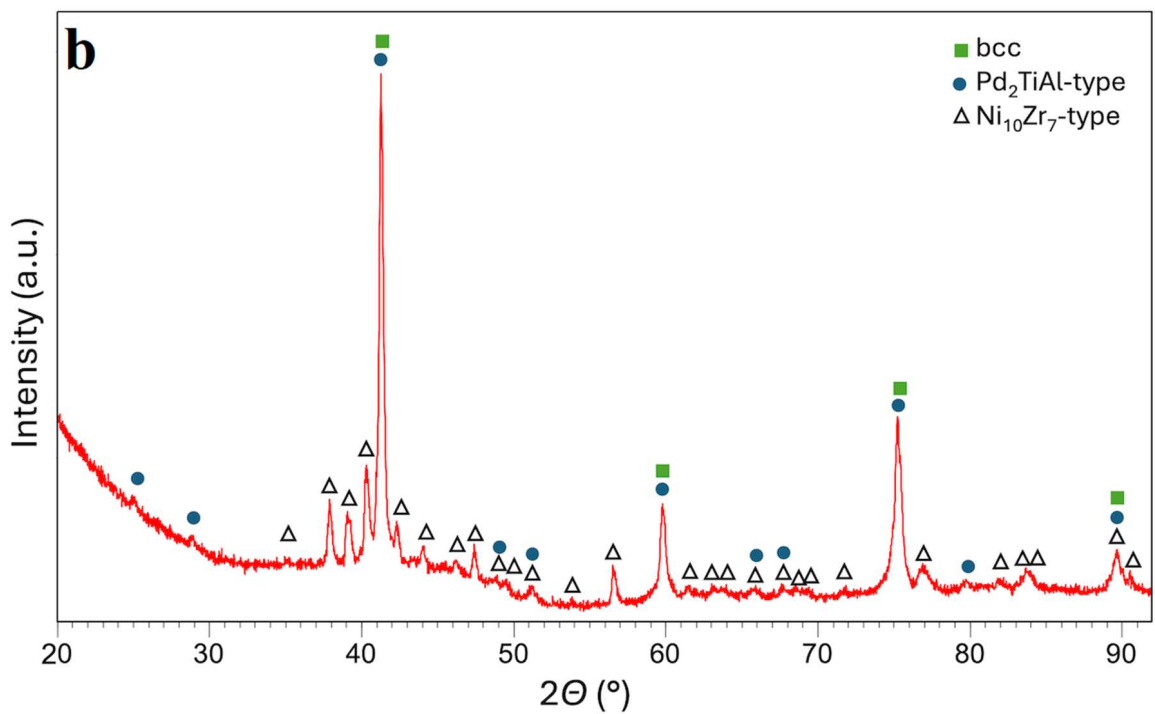

Fig. S1: XRD patterns of the Al$_{0.5}$TiZrPdCuNi alloy in the form of (a) amorphous as-cast ribbon and (b) crystalline cylinder sample (diffraction lines belonging to bcc phase are marked with solid green squares, Pd$_2$TiAl-type phase with solid blue circles and Ni$_{10}$Zr$_7$-like phase with outlined black triangles).

It is worth mentioning that XRD pattern of the amorphous sample which was annealed for 40 min at 600°C (see Fig S2) exhibits broad peaks at the positions similar to those of the sharp crystalline peaks for the cylinder sample indicating the presence of nanometre-sized crystals of the same phases present in the cylinder sample.

Thermal studies and parameters

The Al0.5TiZrPdCuNi ribbons which appeared in X-ray diffraction pattern amorphous (Fig. S1) were further investigated by differential scanning calorimetry (DSC) and thermogravimetric analysis (TGA) using Thermal Analysis DSC-TGA instrument.[26,28] The measurements were performed with a ramp rate of 20 k/min up to 1600 K and the values of thermal parameters were determined by using the TA "Advantage" software.

As seen in Fig. S2 the DSC traces of all ribbons (obtained by melt-spinning of molten alloy from the quartz tubes with orifices of 0.6 and 0.8 mm in diameter, respectively on the surface of copper roller rotating with peripheral velocity of 25 m/s[26,28]) confirmed their amorphous state and provided the values of the thermal parameters which are consistent with these reported in Ref. 34. The values of the glass transition temperature, $T_g$ the onset of the first crystallization temperature, $T_x$ the melting temperature, $T_m$ and the liquidus temperature, $T_l$ were about 769 K, 805 K, 1234 K and 1294 K, respectively. Rather small temperature difference between $T_x$ and $T_g$, $\Delta T_x$=36 K indicates a poor glass forming ability, e.g., Ref. 28. The vertical dashed line at 600 $^0$C indicates the temperature which was used for annealing amorphous ribbons in an unsuccessful attempt to obtain the single-phase bcc crystalline structure.

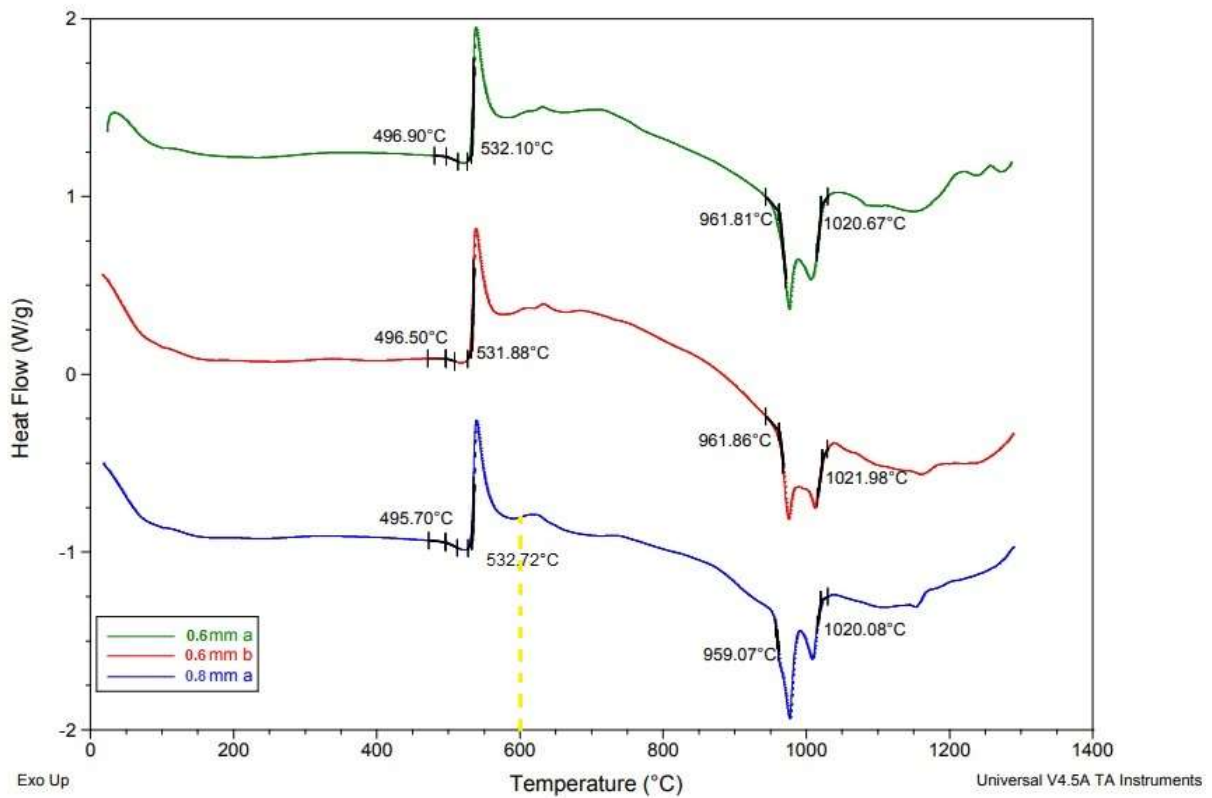

Fig. S2: DSC/DTA traces of Al0.5TiZrPdCuNi ribbons. Note a very good reproducibility of the traces for differently prepared ribbons. The vertical dashed line denotes the temperature used for the crystallization of amorphous ribbons.

Chemical composition and homogeneity

The as-cast ribbons were investigated with the scanning electron microscopy (SEM) using JEOL ISM7600F microscope with the energy dispersive spectroscopy (EDS) capability in order to determine their actual compositions and chemical homogeneity.[26,27,28] The elemental mapping was performed on seven different areas of the ribbon: four areas with a size 30x50 micrometer$^2$ (Fig. S3) and three with larger size of about 300x500 micrometer$^2$.

As illustrated in Fig. S3 Al0.5TiZrPdCuNi samples showed a good chemical homogeneity down to the micrometre scale. The average concentrations of the constituents (in atomic %) obtained from EDS analysis of seven different areas were: 8.6, 16.6, 18.6, 18.8, 21.2 and 16.1 %. Compared to the nominal composition (9.1% of Al and 18.2 % of all other constituents) these from EDS are for most of the constituent elements within the uncertainty of EDS analysis and only the concentrations of Cu and Ni show larger deviations from the nominal ones. However, in the EDS spectra of multi-principal compositionally complex alloys the spectral maxima of some elements overlap, as seen in Fig. S3. A rather strong overlap of some spectral maxima for Cu and Ni in the EDS spectrum (Fig. S3) obviously makes the determination of their individual contributions to overlapping maxima quite uncertain. Note that the actual total concentration of the late transition elements (Pd+Cu+Ni) in our alloy, x(TL) is within the uncertainty of EDS analysis practically the same as the nominal one.

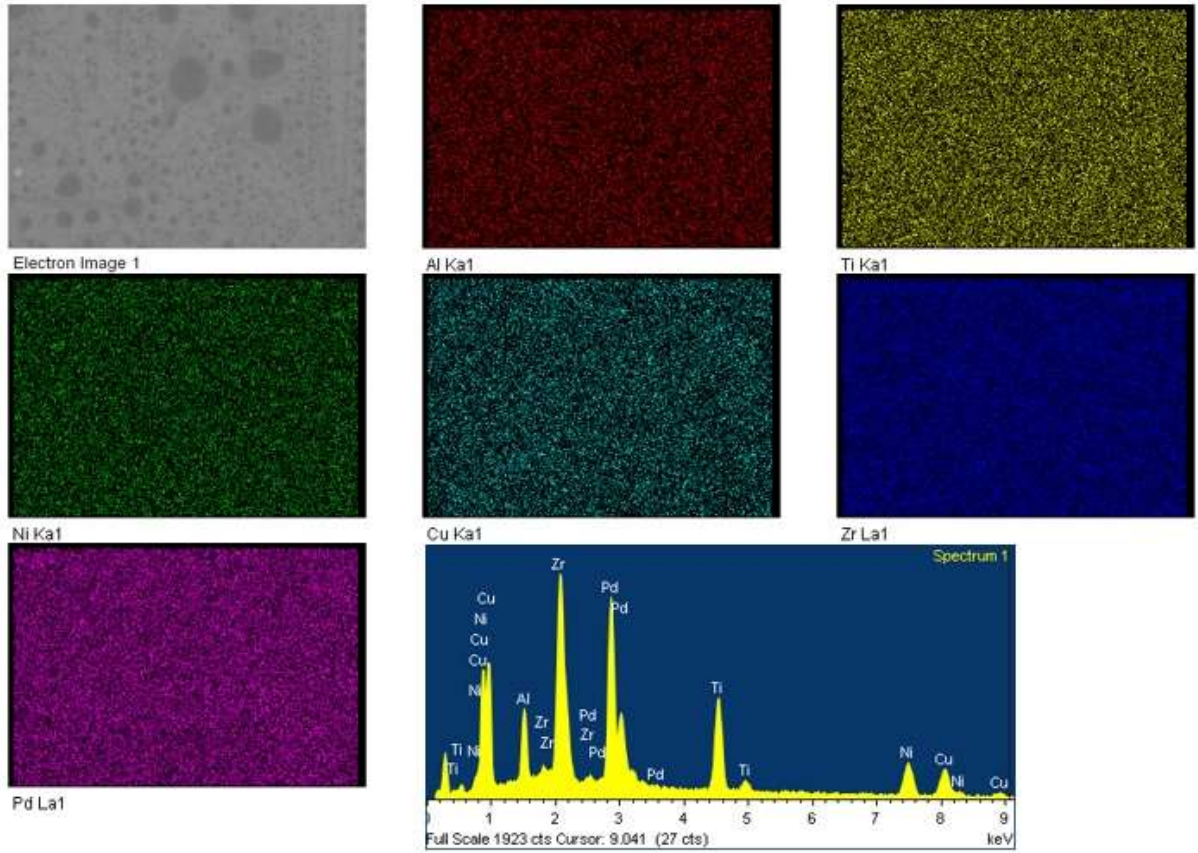

Fig. S3: SEM/EDS images of as-cast Al0.5TiZrPdCuNi ribbon and the corresponding SEM/EDS spectrum.

Acknowledgments

We thank Dr. Vito Despoja for fruitful discussions. E.S., R.R. and K.Z. acknowledge the support of the projects CeNIKS co-financed by the Croatian Government and the European Union through the European Regional Development Fund - Competitiveness and Cohesion Operational Programme (Grant KK.01.1.1.02.0013) and HOI2DEM financed by the Croatian Science Foundation (Grant IP-2022-10-6321). V.M.T. acknowledges support of project Centre for Advanced Laser Techniques (CALT), co-funded by the European Union through the European Regional Development Fund under the Competitiveness and Cohesion Operational Programme (Grant No. KK.01.1.1.05.0001). V.M.T. also acknowledges the support of the project New catalytic materials for the production of green hydrogen – financed by The Environmental Protection and Energy Efficiency Fund
(Croatia), (No. 2023/001520). A.S.F. also acknowledges the support of the project Investigation of the influence of heat treatment on the microhardness of some metallic glasses financed by the Ministry of Science, Higher Education and Youth, Canton Sarajevo, Federation of Bosnia and Herzegovina, Bosnia and Herzegovina (Grant No. 2702114125028/21). D.S. acknowledges support of project Cryogenic Centre at the Institute of Physics — KaCIF co-financed by the Croatian Government and the European Union through the European Regional Development Fund-Competitiveness and Cohesion Operational Programme (Grant No. KK.01.1.1.02.0012).

Author declarations section (conflict of interest, ethics approval, and author contributions)

Conflict of interest

The authors have no conflicts to disclose.

Author contributions

E. Babić: Conceptualization (equal); Writing-original draft; Writing-review & editing (equal). I. A. Figueroa: Investigation (equal); Data curation (equal). V. Mikšić Trontl: Conceptualization (equal); Investigation (equal); Writing-review & editing (equal). P. Pervan: Conceptualization (equal); Investigation (equal); Writing-review & editing (equal); Data curation (equal). I. Pletikosić: Investigation (equal) i Data curation(equal). R Ristić; Investigation (equal); Data curation (equal). A. Salčinović Fetić: Conceptualization (equal); Investigation (equal); Writing-review & editing (equal); Data curation (equal). Ž. Skoko: Investigation (equal); Writing-review & editing (equal). D. Starešinić: Investigation (equal); Data curation (equal). T. Valla: Investigation (equal) i Data curation(equal). K. Zadro: Investigation (equal); Data curation (equal)

Data availability statement

The data that support the findings of this study can be obtained from corresponding authors, E. Babić and A. Salčinović Fetić upon reasonable request.